\documentstyle[12pt]{article}
\def\tr{\mathop{\rm Tr}}

\def\and{\mathop{\rm and}}

\begin{document}
\title{
Ferromagnetism of the 1-D Kondo Lattice Model: \\
A Quantum Monte Carlo Study}
\author{M. Troyer and D. W\"urtz \\
        Interdisziplin\"{a}res Projektzentrum f\"{u}r Supercomputing \\
        and Institut f\"ur Theoretische Physik, ETH Z\"{u}rich,\\
	CH-8092 Z\"urich, Switzerland}
\date{}
\maketitle
\begin{abstract}
The one dimensional Kondo lattice model is investigated using Quantum
Monte Carlo and transfer matrix techniques. In the strong coupling region
ferromagnetic ordering is found even at large band
 fillings.
In the weak coupling region the system shows an RKKY like behavior.
\end{abstract}
%\pagebreak

%\section{Introduction}
In recent years the Kondo lattice model (KLM) has been investigated
to describe the physics of the so called Heavy Fermion systems \cite{review}.
The question was addressed whether this simple model could account for the
rich variety of phases found in Heavy Fermion materials, paramagnetism,
antiferromagnetism as well as superconductivity.
However, as a  typical model of strongly correlated electron systems
it could be analyzed by only few approximate treatments\cite{Read,RiceUeda} and
has still resisted to give clear insight into the various
possible ground state. It would be of particular interest to
understand the phase diagram of the KLM. In this letter we
will tackle this problem by using Quantum Monte Carlo techniques
for the one dimensional KLM.
The results found here partially contradict the phase diagrams given
by Ref.\cite{Fazekas} obtained by variational methods.

The KLM consists of a lattice of $L$ localized spins
($S_{i,f}={1/2};\;i=1,...,L$) coupled to a single band of
 conduction electrons (creation operator
$c_{i,\sigma}^{\dag};\;i=1,...,L;\;\sigma=\uparrow,\downarrow$).
It is described by the Hamiltonian
\begin{eqnarray}
H&=&-t\sum_{\sigma=\uparrow,\downarrow}\sum_{i=1}^L(c^{{\dag}}_{i,\sigma}
c_{i+1,\sigma}+h.c.)\;+\;J\sum_{i=1}^L{\bf S}_{i}\cdot{\bf S}_{i,f},
\\
{\bf S}_i&=&{1\over 2}\left(\matrix{c^{\dag}_{i,\uparrow} & c^{\dag}_{i,
\downarrow}} \right) \vec{\sigma}
 \left(\matrix{c_{i,\uparrow} \cr c_{i,\downarrow}} \right),
\end{eqnarray}
where $\vec{\sigma}$ are the usual Pauli matrices. The first term
denotes the hopping of conduction electrons between nearest neighbor
sites. The second term is an antiferromagnetic ($J>0$) exchange coupling
between localized spins and conduction electrons.
The model can be derived from the periodic Anderson
model in its strong coupling limit \cite{Schrieffer}.

The behavior of the KLM is understood only for some limiting cases.
At half band filling the ground state is a spin singlet \cite{Ueda1}.
The nature of this singlet changes from localized singlet pairs in the
limit $J/t\rightarrow\infty$ to a collective singlet at smaller values of
$J/t$.
A spin gap has been found at intermediate values of $J/t$ using
Quantum Monte Carlo methods \cite{Fye} and for the entire range of the
coupling using exact diagonalization \cite{Tsune}.
On the other hand in the case of only one conduction electron the
ground state of the system is known to be an incompletely saturated
ferromagnet \cite{UedaSigrist}. The same behavior has been found for two
particles and large values of $J/t$ \cite{Sigrist2}.
Questions arising immediately are: what is the ground state of the system
 away from these special cases, where are the boundaries of the
ferromagnetic region? Contrary to expectations from RKKY mechanisms
our results show ferromagnetic ordering
in the strong coupling region even for very large fillings. In the
weak coupling limit an RKKY like behavior is observed at intermediate
temperatures.

%\section{Algorithm}
We have employed a generalization of the standard Quantum Monte Carlo World
Line
Algorithm (WLA) and Quantum Transfer Matrix Algorithm (TMA) \cite{Suzuki}.
The standard checkerboard WLA has to be extended to add the
 localized spins \cite{cluster}.
 There is no hopping in the localized band, therefore
the only local move we have to add is one that can exchange an up (down)
 spin in the conduction band with a down (up) spin in the localized band at
any point of the checkerboard. This move accounts for the
exchange interaction between the two bands. A global move that can
change
the spin of a world line allows for fluctuations in the total magnetization.
The systematic error due to the finite Trotter time step $\Delta\tau$
is controlled by extrapolating from results for $\Delta\tau
t=0.25$ and $\Delta\tau t=0.5$.
The usual zero winding number boundary conditions are used.

Due to the fermionic degrees of freedom there is a sign problem. This
sign problem is most severe at values of $J\sim t$ and near fillings
of $\rho=2/3$, where $\rho=N/L$ and $N$ is the number of conduction
electrons.
For larger or smaller values of $J/t$ and for a filling of $\rho=1/3$
the sign problem is not too severe to make Monte Carlo simulations
impossible.

%\section{Measurements}
To study the magnetic properties of the KLM we have measured the
charge and spin structure factors and susceptibilities. The structure factors
are defined as the Fourier transforms of the real space correlations.
For the conduction band we have for the charge and spin structure
factors:
\begin{eqnarray}
S_{charge}(q)&=&{1\over L}\sum_{j,k}^Le^{iq(j-k)}
(n_{j,\uparrow}+n_{j,\downarrow}) (n_{k,\uparrow}+n_{k,\downarrow}), \\
S_{spin,c}(q)&=&{1\over L}\sum_{j,k}^Le^{iq(j-k)}
(n_{j,\uparrow}-n_{j,\downarrow}) (n_{k,\uparrow}-n_{k,\downarrow})=
{4\over L}\sum_{j,k}^Le^{iq(j-k)}S_j^zS_k^z.
\end{eqnarray}
In the same manner we define the spin structure factor for the localized spins:
\begin{equation}
S_{spin,f}(q)= {4\over L}\sum_{j,k}^Le^{iq(j-k)}S_{j,f}^zS_{k,f}^z.
\end{equation}
The static susceptibility can be calculated as
\begin{equation}\chi(q)={1\over L}\int_0^\beta d\tau\sum_{j,k}^Le^{iq(j-k)}
\tr\left((S_j^z+S_{j,f}^z) e^{-\tau H}(S_k^z+S_{k,f}^z)
 e^{-(\beta-\tau)H}\right),
\end{equation}
where $\beta$ is the inverse temperature $1/T$.

Simulations have been performed on lattices of $L=12,18$ and $24$ sites,
 at band fillings of $\rho={1/3},{2/3}$ and temperatures down to $\beta t=32$.
 Coupling constants
$J/t=0.2,\,0.5,\,1,\,2,\,4$ and $10$ have been investigated. Within the error
bars of our results we cannot see any finite size corrections.  From this
we conclude that the results show properties of the thermodynamic limit.
In the figures we only show the data for the largest reasonable sizes allowed
by the sign problem.
%\section{Results}

In the strong coupling region we observe a tendency toward ferromagnetic
ordering at both fillings. The spin structure factor of the localized spins
 and the susceptibility both show a clear peak at $q=0$ (Fig. 1,2(a)).
 Furthermore the $q=0$ component of the susceptibility is the
 fastest diverging one when lowering the temperature.
We note that this ferromagnetic ordering found at large
band fillings of $\rho=1/3$ and $\rho=2/3$ is not expected from the
RKKY mechanism but is a {\it new} characteristic of the strong coupling region.
Also a slight tendency toward ferromagnetic ordering can be
observed in the spin structure factor of the conduction band (Fig. 2(b)).
Double occupancy of a site is strongly suppressed as it costs energy of
the order of $J$, leading to an effective on-site repulsion.
 Therefore the charge structure factor is essentially that
of spinless fermions, showing a $4k_F$ structure (Fig. 3).

To get more insight into the magnetic properties
we have studied the temperature dependence of the static uniform
susceptibility $\chi(q=0)$ for $J/t=4$ and $\rho={1/3}$ (Fig. 4).
We have simulated systems of $L=6,12,18$ and $24$ sites by the Monte
Carlo method and the $L=6$ site system with the TMA. Special care is
necessary for $L=6$ because the ground state is completely
different depending on the boundary condition.
It has been shown using exact diagonalization that the ground state
is a spin singlet for periodic boundary conditions,
 while it is an incompletely saturated ferromagnet for antiperiodic
boundary conditions \cite{Sigrist2}. At $J/t=4$ the ground state energy
 of the ferromagnetic state is lower than that of the singlet state.
We have calculated $\chi(T)$ both for periodic and antiperiodic
boundary conditions in the canonical ensemble using the TMA.
In the high temperature regime $\chi(T)$ can be obtained for an
infinite size system in the grand canonical ensemble by using the
$M=1$ approximation ($M$ is the Trotter number) in the TMA \cite{Suzuki}.

At high temperatures the susceptibility is the sum of that of a free
conduction electron system and that of free localized spins
 $T\chi={1\over4}(1+\rho-{\rho^2/2})$. When the temperature is
lowered to about $T\approx J$ the conduction electrons start to lock
into singlets with localized spins and the susceptibility reduces to
about the value for the remaining spin degrees of freedom
$T\chi\approx{1\over 4}(1-\rho)$. Lowering the temperature even
further the spins start to order ferromagnetically, leading to an
increase in $T\chi$ and possibly a divergence in an infinite size system.

At small values of $J/t$ a completely different behavior can be
observed. The effective on-site repulsion is smaller, leading to an increase in
the $2k_F$ component of the charge structure factor. The charge
structure factor resembles that of nearly free electrons.
At the same time the $q=0$ component of the conduction electron spin structure
factor and susceptibility is reduced while a cusp emerges at $2k_F$.
In the localized spins a $2k_F$ structure is induced by the
conduction electrons.  This behavior may be called RKKY liquid.
At lower temperatures the $2k_F$ peaks in the spin structure factors become
more pronounced while at the same time the  $q=0$ component is suppressed.
{}From this we conclude that in the temperature regime of our simulations
 ($\beta t\approx32$) the system is dominated by the RKKY interaction.
It does not show a Heavy Fermion behavior,
characterized by a wave vector $k_F'=(\rho+1)\pi/2$, including both
conduction and localized electrons.

A problem arising in the small $J/t$ region is that the effective
coupling gets very small. Although the temperature of the Monte Carlo
simulations
is well below the Fermi temperature, it is still orders of magnitude
above the single impurity Kondo temperature $T_K$.
 $T_K\approx 10^{-5}\; (10^{-9})$ at $J/t=0.5$ and
 $\rho={1/3}\;({2/3})$.
 This makes it hard to get information on the ground state
 of the system from Monte Carlo simulations.

In figure 5 we summarize the results of our Monte Carlo
simulations. At large values of $J/t$ we find
ferromagnetic ordering. There the $q=0$ component of the
susceptibilities is the dominant and fastest diverging one.
{}From our result we conclude that the ferromagnetic state
is stable for a much wider region of the phase diagram than
suggested in Ref.\cite{Fazekas}.
%Although our results indicate ferromagnetic behavior it is not possible to
%prove rigorously
%by Monte Carlo simulations that the KLM has a ferromagnetic ground state.
In the small $J/t$ region the
system shows an RKKY liquid behavior, where the $q=2k_F$
component of the susceptibilities and spin structure factors is the
dominant one. At $J/t=1$ and $\rho=1/3$ the
system shows a transitional behavior. Probably this point is near the phase
boundary \cite{disclaimer}.

The calculations were done on the Cray Y-MP/364 of ETH Z\"urich and
on the NEC SX-3/22 of the Centro Svizzero di Calcolo Scientifico CSCS Manno.
 The work was supported by the Swiss National Science Foundation under
grant number SNF 21-27894.89 and by an internal grant of ETH
Z\"urich.  We want to thank T.M. Rice, K. Ueda, H. Tsunetsugu and M. Sigrist
for
stimulating discussions.

\begin{sloppypar}
\section*{Figure captions}
\newcounter{bean}
\begin{list}%
{Fig. \arabic{bean}}{\usecounter{bean}
                   \setlength{\rightmargin}{\leftmargin}}
\item
Static uniform susceptibility for the Kondo lattice model at $\rho=1/3$ and
$J/t$ ranging from $0.5$ to $4$. The temperature is $\beta t=24$.
At this filling $2k_F=\pi/3$.
\item
Spin structure factors for (a)  the localized spins and (b) the conduction band
for the same parameters as in Fig. 1.
\item
Charge structure factors for the same parameters as in Fig. 1.
\item
Temperature dependence of the static uniform susceptibility for
a filling of $\rho=1/3$. The solid lines show the $M=1$ approximation
for high temperatures and the transfer matrix (TMA) results for
periodic (PBC) as well as antiperiodic (APBC) boundary conditions.
Monte Carlo data for $L=6$ and $L=12$ sites and zero winding
boundary conditions are included. Due to the sign problem the results
for larger lattices have much larger error bars. They show no qualitatively
different behavior.
\item
Phase diagram of the Kondo Lattice Model. The Monte Carlo simulations have been
performed at the points shown here. The different symbols denote points with
ferromagnetic, transitional and RKKY liquid behavior.
\end{list}

\end{sloppypar}


\begin{thebibliography}{99}
\bibitem{review} For a review, P.A. Lee,~T.M.~Rice, J.W.~Serene,
	L.J.~Sham, and J.W.~Wilkins, Comments Condens.~Matt.~Phys.
	{\bf 12}, 99 (1986).

\bibitem{Read} N.~Read, D.M.~Newns, and S.~Doniach, Phys.~Rev.
	{\bf B 30}, 3841 (1984); P.~Coleman, Phys.~Rev. {\bf B 29},
	3035  (1984).

\bibitem{RiceUeda} T.M.~Rice and K.~Ueda, Phys.~Rev.~Lett. {\bf 55},
	995  (1985); Phys.~Rev. {\bf B 34}, 6420 (1986).

\bibitem{Fazekas} P.~Fazekas and E.~M\"uller-Hartmann, Z. Phys. B - Condensed
Matter {\bf 85}, 285 (1991)

\bibitem{Schrieffer} J.R.~Schrieffer and P.A.~Wolff, Phys.~Rev.~{\bf 149},
	491 (1966).

\bibitem{Ueda1} K.~Ueda, H.~Tsunetsugu, and M.~Sigrist, Phys.~Rev.~Lett.
	{\bf 68}, 1030 (1992).

\bibitem{Fye} R.M.~Fye and D.J.~Scalapino, Phys.~Rev.~Lett. {\bf 65},
	3177 (1990); Phys.~Rev.~{\bf B~44}, 7486 (1991).

\bibitem{Tsune}  H.~Tsunetsugu, Y.~Hatsugai, K.~Ueda, and
	M.~Sigrist, to be published.

\bibitem{UedaSigrist} K.~Yamamoto and K.~Ueda, J.~Phys.~Soc.~Jap.
	{\bf 59}, 3284 (1990);
	M.~Sigrist, H.~Tsunetsugu, and K.~Ueda, Phys.~Rev.~Lett.
	{\bf 67}, 2211 (1991).

\bibitem{Sigrist2}
	M.~Sigrist, K.~Ueda, and H.~Tsunetsugu, to be published

\bibitem{Suzuki} M.~Suzuki (editor), Quantum Monte Carlo Methods, Springer
        Series in Solid-State Sciences 74 (Springer Verlag, Berlin, 1987)
        and references given therein.

\bibitem{cluster} M.~Troyer, Helvetica Physica Acta~{\bf 64}, 700 (1991).
\bibitem{disclaimer}
An open question is whether the
behavior we can observe using Monte Carlo techniques is the ground
state behavior or there is a crossover to a different state
at lower temperatures. Further investigations are necessary.

\end{thebibliography}
\end{document}